\documentclass[conference]{IEEEtran}
\usepackage{graphicx}
\usepackage{subfig}
\usepackage{stfloats}
\usepackage{array}

\def\PANDA{$\overline{\mbox{P}}${ANDA }}%

\begin{document}
\title{The Straw Tube Trackers of the \PANDA Experiment}
\author{\IEEEauthorblockN{P.~Gianotti\IEEEauthorrefmark{1}, V.~Lucherini\IEEEauthorrefmark{1}, E.~Pace\IEEEauthorrefmark{1},
G.~L.~Boca\IEEEauthorrefmark{2}, S.~Costanza\IEEEauthorrefmark{2}, P.~Genova\IEEEauthorrefmark{2}, L.~Lavezzi\IEEEauthorrefmark{2}, 
P.~Montanga\IEEEauthorrefmark{2}, \\A.~Rotondi\IEEEauthorrefmark{2}, M.~Bragadireanu\IEEEauthorrefmark{3}, 
M.~E.~Vasile\IEEEauthorrefmark{3}, D.~Pietreanu\IEEEauthorrefmark{3}, J.~Biernat\IEEEauthorrefmark{4}, S.~Jowzaee\IEEEauthorrefmark{4}, 
G.~Korcyl\IEEEauthorrefmark{4}, M.~Palka\IEEEauthorrefmark{4}, \\ P.~Salabura\IEEEauthorrefmark{4}, J.~Smyrski\IEEEauthorrefmark{4},
T. Fiutowski\IEEEauthorrefmark{5}, M.~Idzik\IEEEauthorrefmark{5}, D.~Przyborowski\IEEEauthorrefmark{5}, K.~Korcyl\IEEEauthorrefmark{6}, 
P.~Kulessa\IEEEauthorrefmark{6}, K.~Pysz\IEEEauthorrefmark{6},\\ S.~Dobbs~\IEEEauthorrefmark{7}, A.~Tomaradze\IEEEauthorrefmark{7}, 
D.~Bettoni\IEEEauthorrefmark{8}, E.~Fioravanti\IEEEauthorrefmark{8}, I.~Garzia\IEEEauthorrefmark{8}, M.~Savri\`e\IEEEauthorrefmark{8},
V.~Kozlov\IEEEauthorrefmark{9}, M.~Mertens\IEEEauthorrefmark{9},\\ H.~Ohm\IEEEauthorrefmark{9}, S.~Orfanitski\IEEEauthorrefmark{9}, J.~Ritman\IEEEauthorrefmark{9}, 
V.~Serdyuk\IEEEauthorrefmark{9}, P.~Wintz\IEEEauthorrefmark{9}, S.~Spataro\IEEEauthorrefmark{10}}

\IEEEauthorblockA{\IEEEauthorrefmark{1}INFN, Lab. Naz. Frascati, via E. Fermi 40, 00044 Frascati, Italy}
\IEEEauthorblockA{\IEEEauthorrefmark{2}Univ. and INFN Pavia, via Bassi 6, I-27100 Pavia, Italy}
\IEEEauthorblockA{\IEEEauthorrefmark{3}IFIN H.H., Str. Reactorului 30, Bucharest - Magurele, Romania}
\IEEEauthorblockA{\IEEEauthorrefmark{4}Inst. of Physics Jagiellonian Univ., ul. Reymonta 4, 30-059 Krakow, Poland}
\IEEEauthorblockA{\IEEEauthorrefmark{5}Detector Dev. Lab.,  AGH Univ. of Science and Technology, 30 Mickiewicza Av., 30-059 Krakow, Poland}
\IEEEauthorblockA{\IEEEauthorrefmark{6}Inst. of Nuclear Physics,  PAN, ul. Radzikowskiego 152, 31-342 Krakow, Poland}
\IEEEauthorblockA{\IEEEauthorrefmark{7}Dep. of Physics \& Astronomy Northwestern Univ., 2145 Sheridan Road, Evanston, IL 60208-3112, USA}
\IEEEauthorblockA{\IEEEauthorrefmark{8}Univ. and INFN Ferrara,  via Saragat 1, 44122 Ferrara, Italy}
\IEEEauthorblockA{\IEEEauthorrefmark{9}IKP Forschungszentrum J\"ulich,  Geb. 07.1 Wilhelm-Johnen-Strasse, 52425 J\"ulich, Germany}
\IEEEauthorblockA{\IEEEauthorrefmark{10}Univ. and INFN Torino,  via P. Giuria 1, 10125 Torino, Italy}
}

\IEEEspecialpapernotice{(Invited Paper)}
\maketitle
\thispagestyle{empty}

\begin{abstract}
The \PANDA experiment will be built at the FAIR facility at Darmstadt (Germany) to perform accurate tests of the 
strong interaction through $\bar pp$ and $\bar pA$ annihilation's studies. 
To track charged particles, two systems consisting of a set of planar, closed-packed, self-supporting straw tube 
layers are under construction.  
The \PANDA straw tubes will have also unique characteristics in term of material budget and performance. They consist 
of very thin mylar-aluminized cathodes which are made self-supporting by means of the operation gas-mixture over-pressure. 
This solution allows to reduce at maximum the weight of the mechanical support frame and hence the detector material 
budget. The \PANDA straw tube central tracker will not only reconstruct charged particle trajectories, but also 
will help in low momentum ($<$ 1 GeV) particle identification via dE/dx measurements. 
This is a quite new approach that \PANDA tracking group has first tested with detailed Monte Carlo simulations, and 
then with experimental tests of detector prototypes. This paper addresses the design issues of the \PANDA 
straw tube trackers and the performance obtained in prototype tests.
\end{abstract}


\section{Introduction}
\begin{figure}[h!]
\includegraphics[width=\columnwidth]{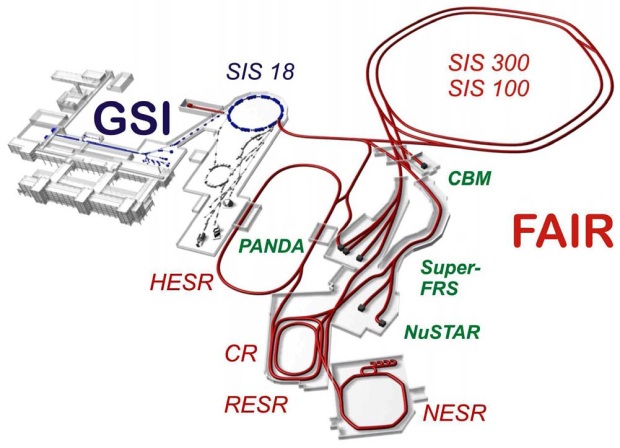}
\caption{Layout of the FAIR facility.}
\label{fig:fair}
\end{figure}

%
%
%
%
FAIR \cite{FAIR} is a new international accelerator facility for the research with antiprotons and ions, under construction near Darmstadt in Hesse, Germany. It is a major upgrade of the existing GSI Laboratory whose accelerator complex will serve as a FAIR injection stage.
The central part of FAIR is a synchrotron complex providing intense pulsed ion beams (from $p$ to $U$). The layout of the FAIR facility 
is depicted in Fig. \ref{fig:fair}.

The antiprotons will be available into the High Energy Storage Ring (HESR), a slow ramping accelerator hosting a single internal interaction point enclosed within the \PANDA  detector \cite{PANDA}.
\PANDA will investigate QCD in the charmonium mass range, and other aspects of particle and nuclear physics via $\bar p-p$
and $\bar p-N$ annihilations \cite{PB}.
\par
\begin{figure}[h!]
\includegraphics[width=\columnwidth]{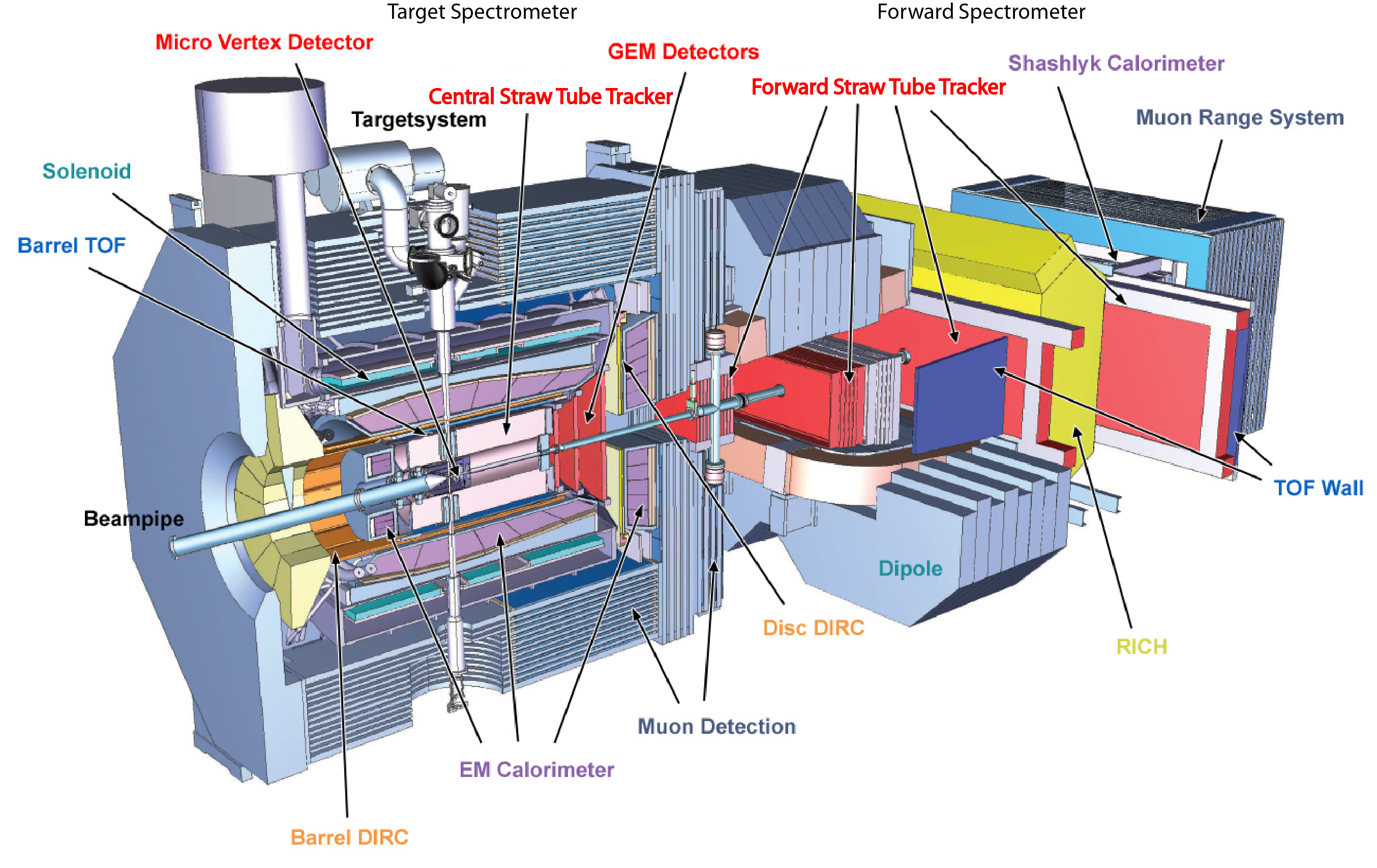}
\caption{Drawing of the \PANDA apparatus. The antiproton beam is entering from the left side. }
\label{fig:panda}
\end{figure}

\PANDA will be a fixed target experiment with a Target Spectrometer (TS), surrounding
the interaction point, and a Forward Spectrometer (FS) to close the acceptance at forward angles (see Fig. \ref{fig:panda}).
\par
Tracking charged particles is one of the essential tasks of the \PANDA experiment, hence 
different tracking devices are under construction for the two spectrometers and among them two straw tube trackers: 
the first, located in the TS, is a cylindrical device; the second, in the FS, consists of a set of planar chambers.
A new technique, based on the use of straw tubes operated at over-pressure has been chosen. This allows to construct 
self-supporting modules avoiding heavy mechanical frames. 

The measurement of the particle's energy loss is also 
implemented in the central tracker. For momenta up to 1 GeV/$c$ an energy resolution $\sigma_E <$10\% can be obtained. 
This will provide extra information, helpful for the particle identification process.

\section{\PANDA straw tubes}
\begin{figure}[h!]
\begin{center}
\includegraphics[width=\columnwidth]{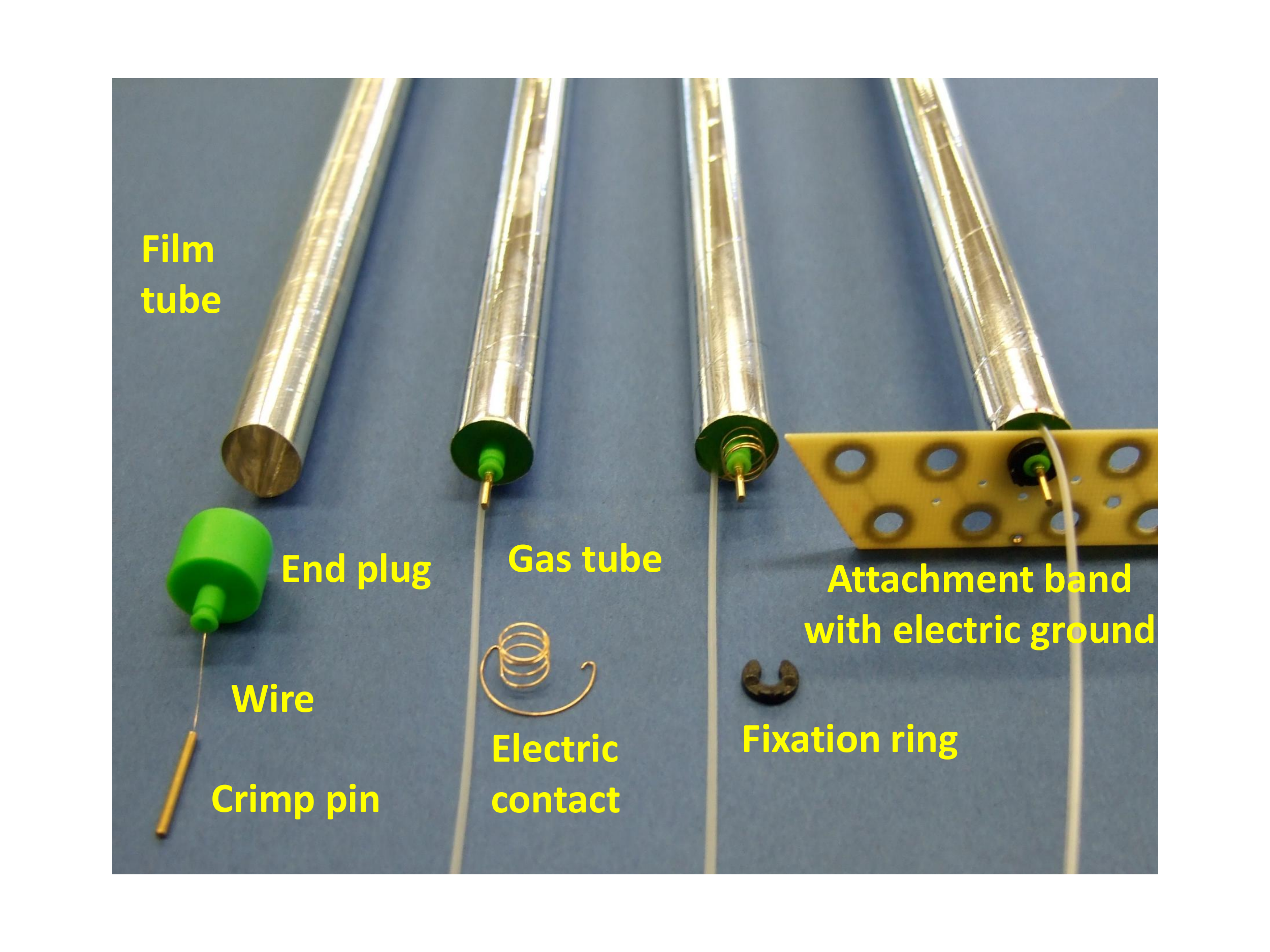}
\caption{Straw tube components (see text for more details).}
\label{fig:stt}
\end{center}
\end{figure}

The two \PANDA straw tube detectors will both use self-supporting
straw tube modules made of similar straws of 10 mm diameter. The cathode material is a thin
aluminized mylar film (thickness 27 $\mu$m) with a gold-plated tungsten-rhenium wire,
of 20 $\mu$m diameter, as anode. The anode wire is stretched by a weight of 50 g and
crimped in copper, gold-plated pins.
Light end-plugs, of ABS material with a wall thickness of 0.5 mm, close
the tubes at both ends and provide the pin and the gas pipe housing. They are glued to
the mylar film leaving a small 1.5 mm film overlap on both ends. There, a gold-plated
copper-beryllium spring is placed with a double function: to provide cathode grounding
and to compensate the tube elongation induced by the over-pressure of the gas mixture 
(see Fig. \ref{fig:stt}).
The total weight of a fully assembled straw tube is only 2.5 g.

The application on each wire of a tension, in the case of the 4636 straws of the central tracker, 
results in a force equivalent to $\sim$ 230\,kg which must be maintained.
Usually, this is done by supporting the straw tubes with a strong and massive 
frame or by adding reinforcement structures. This inevitably increases the
detector radiation length and it is not acceptable for the \PANDA central tracker.
Therefore, it has been decided to adopt a technique based on self-supporting straw
layers developed for the COSY-TOF straw tracker \cite{COSY-TOF}.
The straw tube wires are stretched by means of an over-pressure of 1\,bar of the gas mixture. 
Then the tubes are close-packed and glued together to planar multi-layers on a reference table which defines a
precise distance of 10.1\,mm between adjacent tube's centers. At the gas over-pressure
of 1\,bar a double-layer not only maintain the nominal wire's tension, but also become self-supporting.
\par
\begin{figure}[h!]
\begin{center}
\includegraphics[width=2.5in]{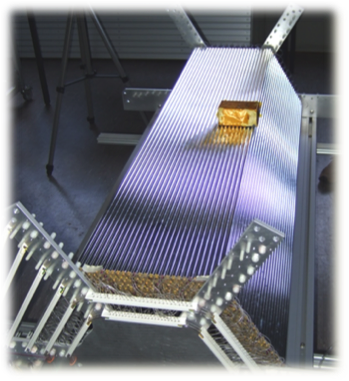}
\caption{Pressurized, close-packed straw layers show strong rigidity
as demonstrated here by a 3 kg Pb-brick.}
\label{fig:brick}
\end{center}
\end{figure}

Fig. \ref{fig:brick} shows a pressurized straw multi-layer. The system is
holding the weight of a Pb brick of 3 kg.

\begin{figure}[!h]
\begin{center}
\includegraphics[width=\columnwidth]{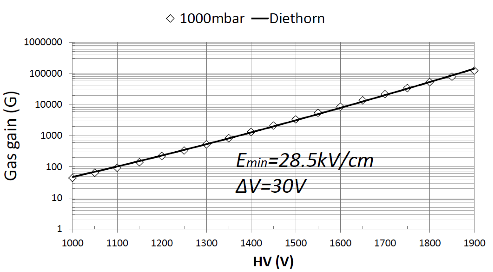}
\caption{Gas gain measurements performed in a straw tube prototype as a function of the applied high voltage. 
The points represent the experimental measurements while the line is the fit done with a Diethorn's function.
Best fit parameters are also indicated.}
\label{fig:gasgain}
\end{center}
\end{figure}

The gas mixture used for both \PANDA straw tube trackers is Argon based with 10\% CO$_2$ as quencher.
This gas mixture is known as being one of the best for high-rate hadronic environments
due to the absence of polymer reactions of the components once
there is a clean gas environment including all materials and parts of the detector
and gas supply system.
The high voltage is set to have a gas gain of about 10$^5$ in order to warrant long term operation.
Fig. \ref{fig:gasgain} shows the results of the gas gain measurements performed on a detector prototype \cite{jinst}.

\subsection{The central tracker}
\begin{figure*}[!b]
\centering
\subfloat[]{\includegraphics[width=2.5in]{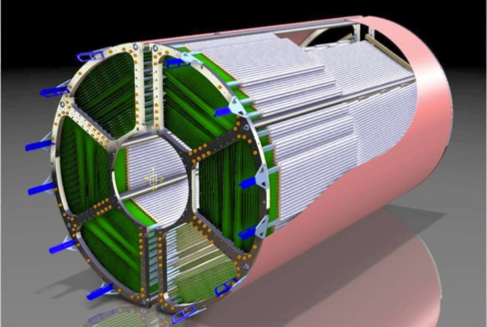}}
\hfil
\subfloat[]{\includegraphics[width=2.5in]{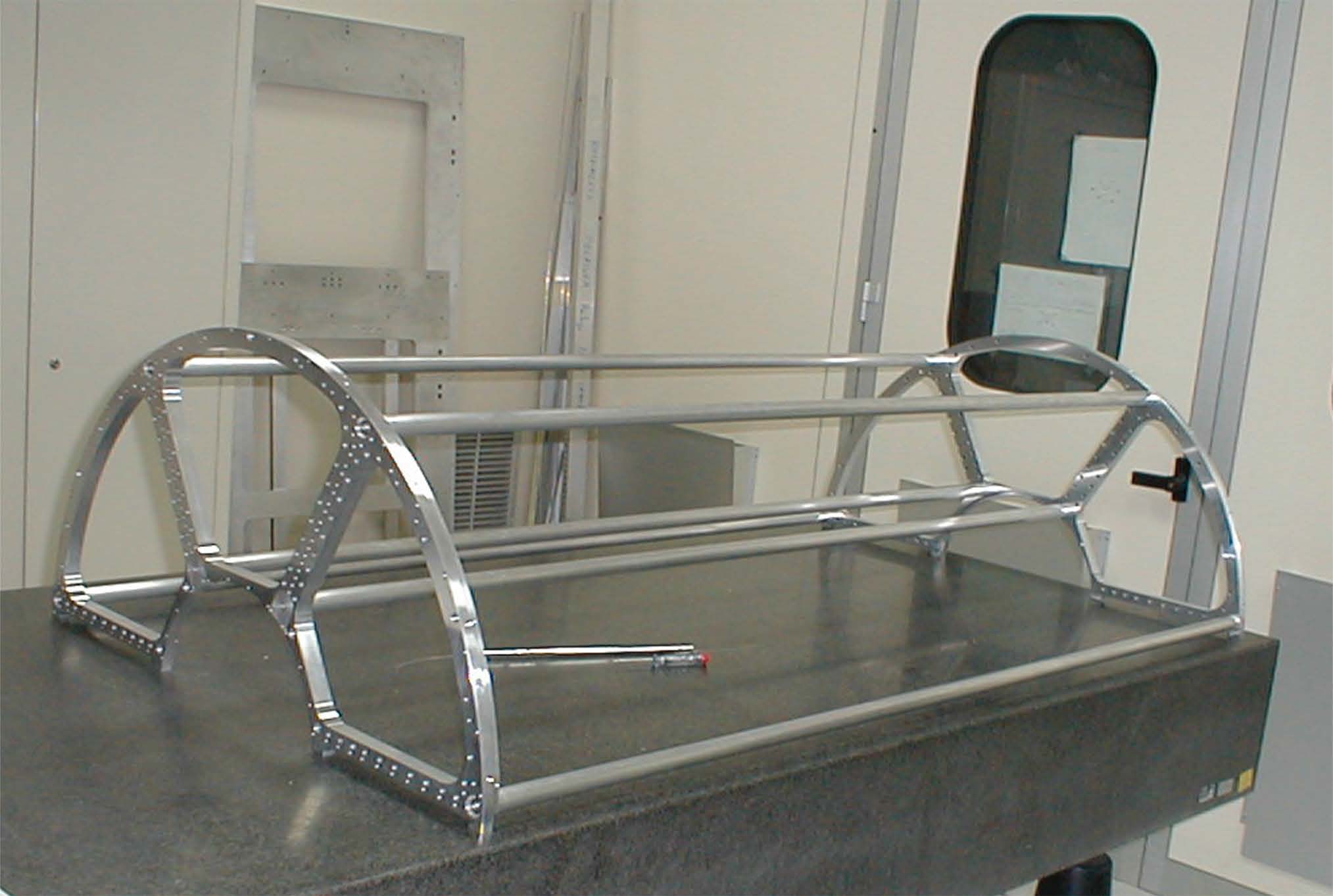}}
\caption{a) CAD drawing of the central straw tube tracker. b) Prototype of the frame holding one semi-chamber (right).}
\label{fig:layout}
\end{figure*}

The \PANDA central straw tube tracker occupies a cylindrical volume with an inner radius of 150 mm and an outer one of 418 mm.
Due to the presence of the target pipe, in the $x,y$ plane this volume is divided in two halves, with a gap of
42 mm in between. Along $z$, the allowed space is 1500 mm, plus 150 mm in the upstream region for electronics, gas supplies,
and other services.
To fill up this volume, it has been decided to use planar layers mounted in a hexagonal arrangement.
The detector layout consists of layers parallel to the beam axis, interleaved with skewed layers, with an angle,
with respect to the beam axis, of about $\pm 3^o$, to allow a 3D reconstruction of particle's trajectories.
In total the detector consists of 4636 straws divided in two identical semi-chambers held by two light mechanical frames
of only 8.2 kg weight in case of Aluminum.
The overall detector will result in a material budget of 1.2\% of a radiation length.
Fig. \ref{fig:layout} shows a CAD drawing of the proposed arrangement and a prototype of the support frame of one semi-chamber.
A complete description of this detector can be found in ref. \cite{tdr}.

\subsection{The forward tracker}
The particle trajectories in the Forward Spectrometer will be measured by 
three pairs of planar tracking stations. These are made of straw tubes similar to those used in the central tracker.
The first pair will be placed in front, the second within and the third behind a dipole magnet providing a
field integral up to 2\,Tm (see Fig. \ref{fig:panda}) that will be properly scaled 
to match the momentum of the antiproton beam.
The angular acceptance of this system, defined by the aperture of the dipole magnet, is equal to $\pm 10^{\circ}$
in the horizontal plane and $\pm 5^{\circ}$ in the vertical one.
\par
Each of the six tracking stations of the forward tracker consists of four double-layers: 
the first and the fourth one contain vertical straws and the two intermediate double-layers 
- the second and the third one - contain straws inclined at $+5^{\circ}$ and $-5^{\circ}$, respectively.
\par
\begin{figure}
\begin{center}
\includegraphics[width=3.5in]{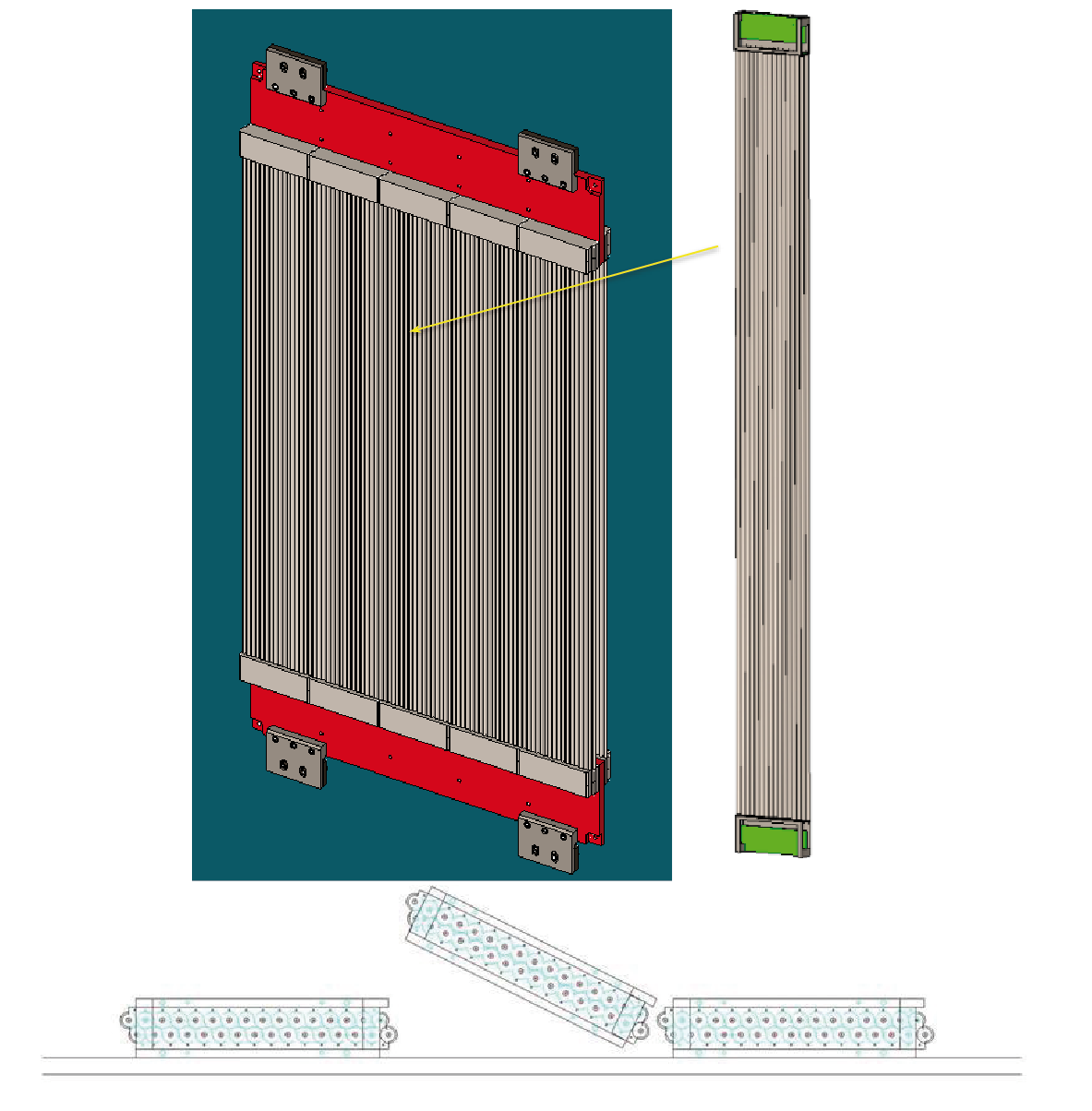}
\caption{(top) A group of five modules of the \PANDA forward straw tube system mounted on the mechanical support frame.
(bottom) Schematic drawing explaining the mounting scheme of a single double-layer module.
The proposed solution allows to install individual modules without removing the two neighborings.}
\label{fig:ft}
\end{center}
\end{figure}

The detection planes are built of separate modules, each containing 32 straws, arranged in two layers,
as shown in Fig.~\ref{fig:ft}.
Each module is equipped with gas supply and its own front end electronic card.
A module can be mounted and dismounted from the support frame without removing the neighborings
(see lower part of Fig.~\ref{fig:ft}).

The supports of the tracking stations, mounted before and after the dipole magnet, consist
of rectangular chassis made of steel profiles and of two pairs of c-shaped frames called drawers
which are connected to the chassis with telescopic rails.
One pair, including the left and the right drawers, is used for supporting two double layers of straws.
The drawer can be moved sidewards with respect to the beam pipe allowing an easy
access to the straw modules.
For the tracking stations mounted inside the dipole magnet gap, the concept of supports with drawers
can not be applied due to lack of space.
Instead, the modules are mounted on closed rectangular frames made of aluminum.
\par
\begin{figure*}[!b]
\begin{center}
\subfloat[]{\includegraphics[scale=0.5]{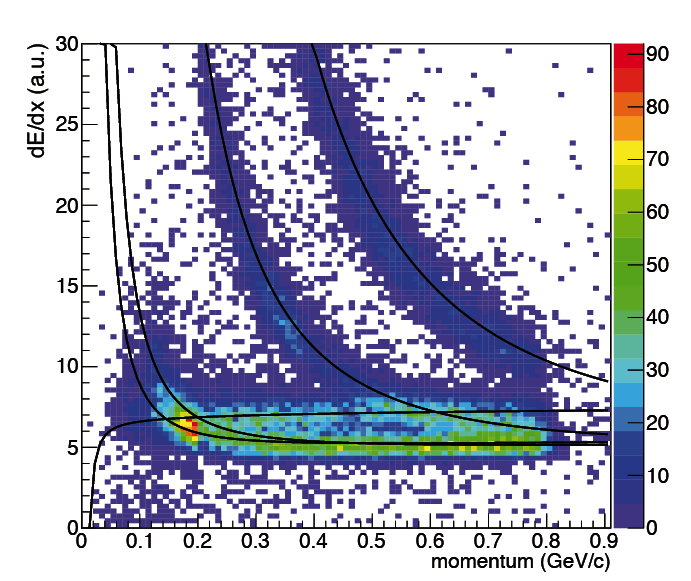}
\label{fig:dedx:sim1}}
\subfloat[]{\includegraphics[scale=0.45]{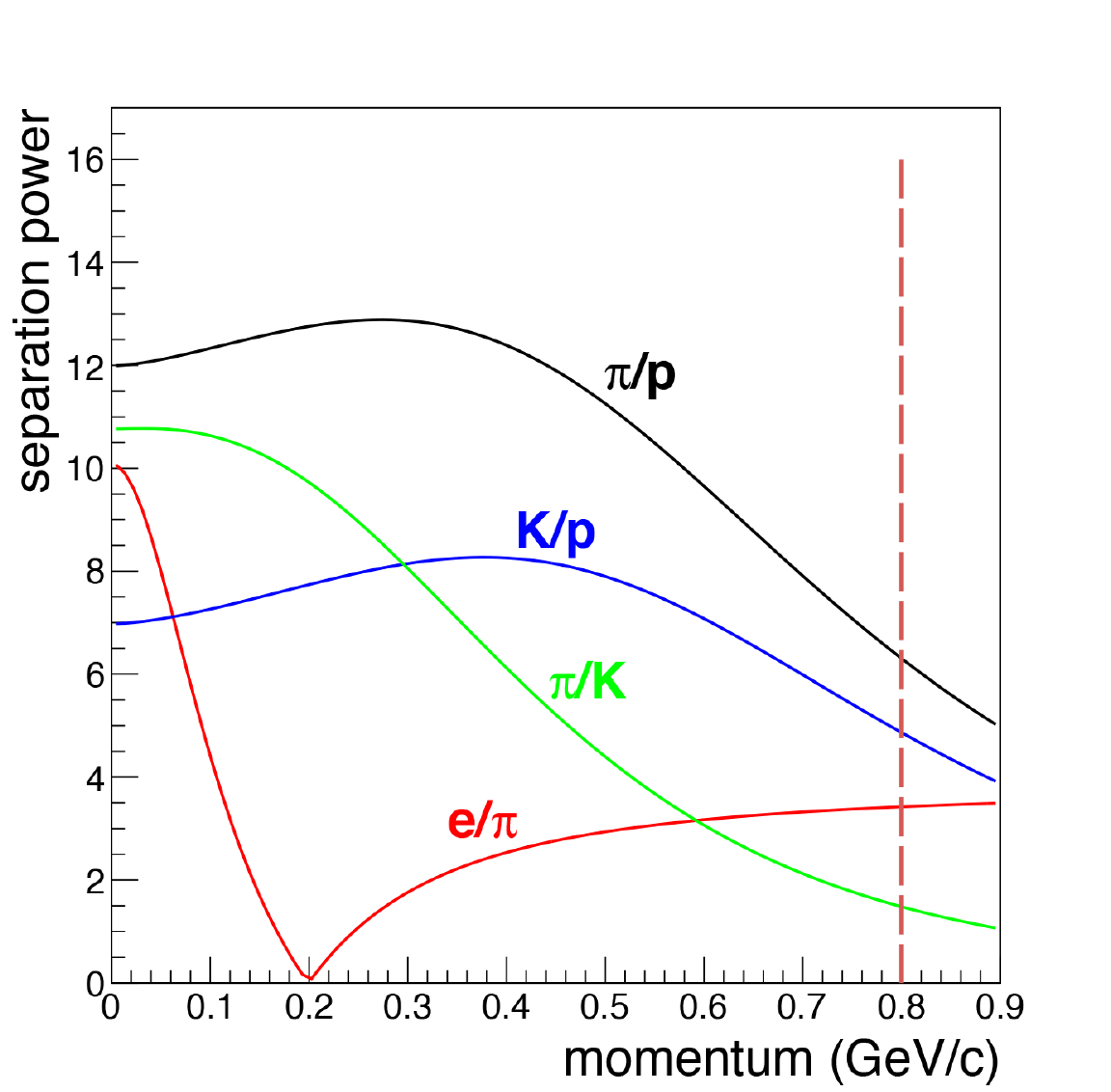}
\label{fig:dedx:sim2}}
\caption{a) Simulated distribution of dE/dx vs momentum for different particles. b) Separation power for different
particle pairs.}
\end{center}
\end{figure*}
The active area of the first tracking station, mounted 3\,m
downstream of the target, is about 0.7\,m high and 1.3\,m wide.
The last  tracking station, being the largest one, is placed 7.5\,m downstream the target
and its active area is 1.5\,m high and 6\,m wide. 
The whole system comprises about 13000 straw tubes. 
In the central part of each double layer, single straws are replaced with shorter ones 
in order to leave a free space for the beam pipe passage. This solution has been already used within the COSY-TOF
experiment for the same purpose. Two adjacent shorter straws, on the cut side,
do not have the copper-beryllium spring and the gas pipe coming out from the first is used as the gas in-let for the second.

\section{Prototype tests}

The straw tube tracker of the \PANDA Target Spectrometer has to help in the process of particle identification
for momenta below 1 GeV/$c$. This is something not routinely done in other similar detectors,
therefore this possibility has been deeply studied with Monte Carlo simulations and experimental tests.
\par
Fig. \ref{fig:dedx:sim1} shows the distribution of the simulated specific energy losses for different particles,
plotted versus the momentum.

The particle's path has been reconstructed via the simulated drift time and by the dip angle resulting from the fit.
The dE/dx has been calculated summing up energy deposits measured by the straws crossed by the particle, and applying a
truncated mean at 30\% in order to cut out the high dE/dx tails.
The separation power between two particles species is defined as the distance between the centres of the two specific energy loss
bands.
Particle's identification is feasible for momenta below 0.8 GeV/$c$, if the resolution on the energy-loss measurement is of about 
10\%, see Fig. \ref{fig:dedx:sim2}.
\par
\begin{figure}[!h]
\begin{center}
\includegraphics[scale=0.5]{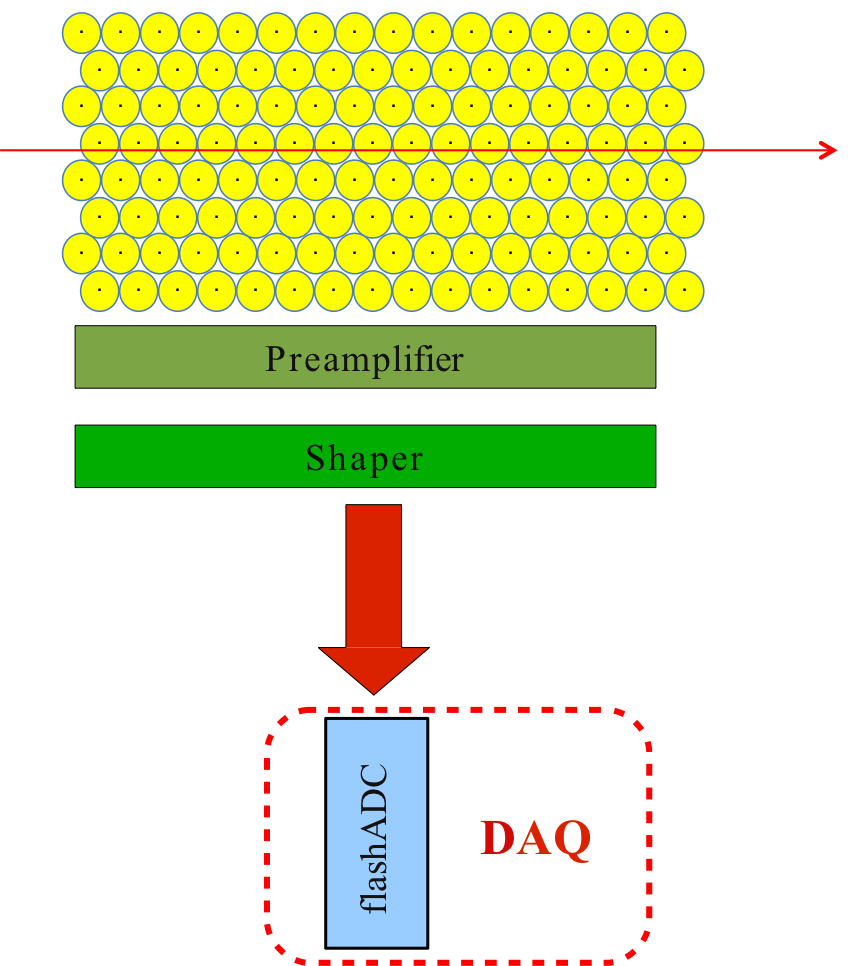}
\caption{Experimental setup of the straw tube prototype test. The red line indicates the proton beam. 
Straw tube signals have been processed with a discrete components electronic chain, and then digitized with a 240 MHz flashADC.}
\label{fig:exp_setup}
\end{center}
\end{figure}
\begin{figure}
\begin{center}
\includegraphics[width=\columnwidth]{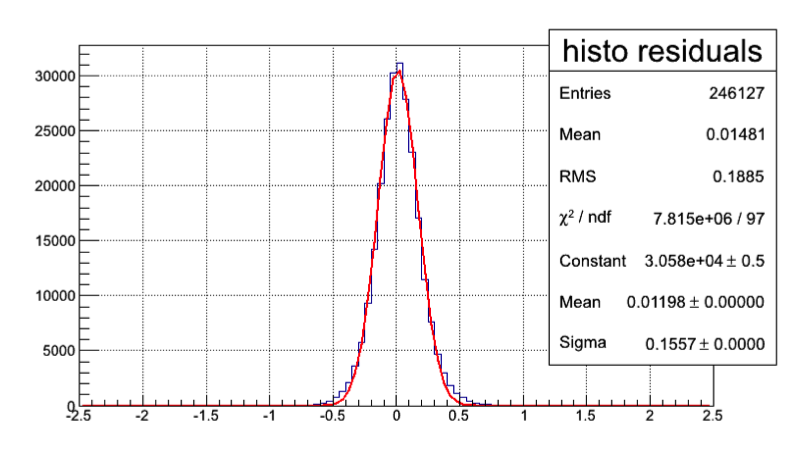}
\caption{Distribution of the spatial residuals obtained for the test beam of the straw tube tracker prototype. A spatial resolution of about 
150 $\mu$m has been achieved.}
\label{fig:res}
\end{center}
\end{figure}

To test experimentally these results, a detector prototype of 128 straw tubes arranged in four double-layers of 32
straws each has been exposed to an extracted proton beam of different momenta, at the COSY synchrotron of the J\"ulich 
Research Center \cite{cosy}.
The arrangement of the test setup is shown in Fig. \ref{fig:exp_setup}.
The spatial resolution obtained during this test is $\sim$ 150 $\mu$m as shown in Fig. \ref{fig:res}.

\begin{figure}
\begin{center}
\includegraphics[width=\columnwidth]{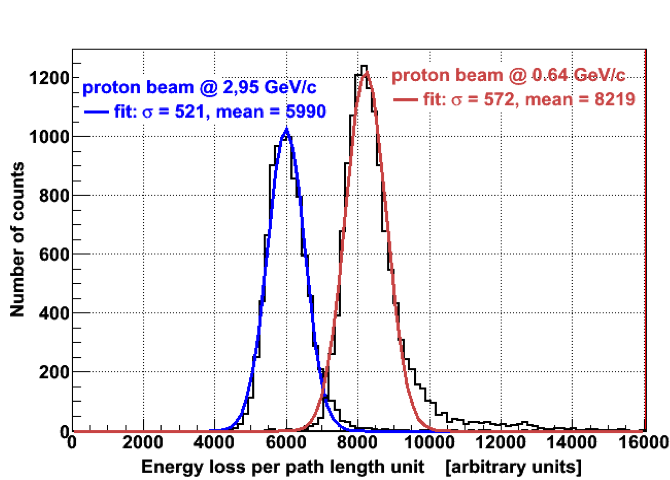}
\caption{dE/dx distributions for protons of 2.95 GeV/$c$ and 0.64 GeV/$c$ with Gaussian fits. Truncated mean of 30\% is applied to all 
histograms.}
\label{fig:dedx:exp}
\end{center}
\end{figure}

To evaluate the specific energy loss of the particles, straw tube signals have been processed with a 240 MHz flashADC.
For each event, the amplified signals from the fired straws have been summed up to build the energy loss distribution and,
as for the simulations, the truncated mean technique has been applied to cut out the high energy tails.
The truncated distributions of the energies have been then divided by the appropriate reconstructed track lengths.
The results, for 2 different proton momenta, are shown in Fig. \ref{fig:dedx:exp}. The Gaussian fits are superimposed and the fit parameters are given in the figures.
The best energy resolution achieved is $\sigma_{dE/dx}= 7 \pm 1 \%$ for proton of 0.64 GeV/$c$ momentum.
\par
\begin{figure}
\begin{center}
\includegraphics[width=\columnwidth]{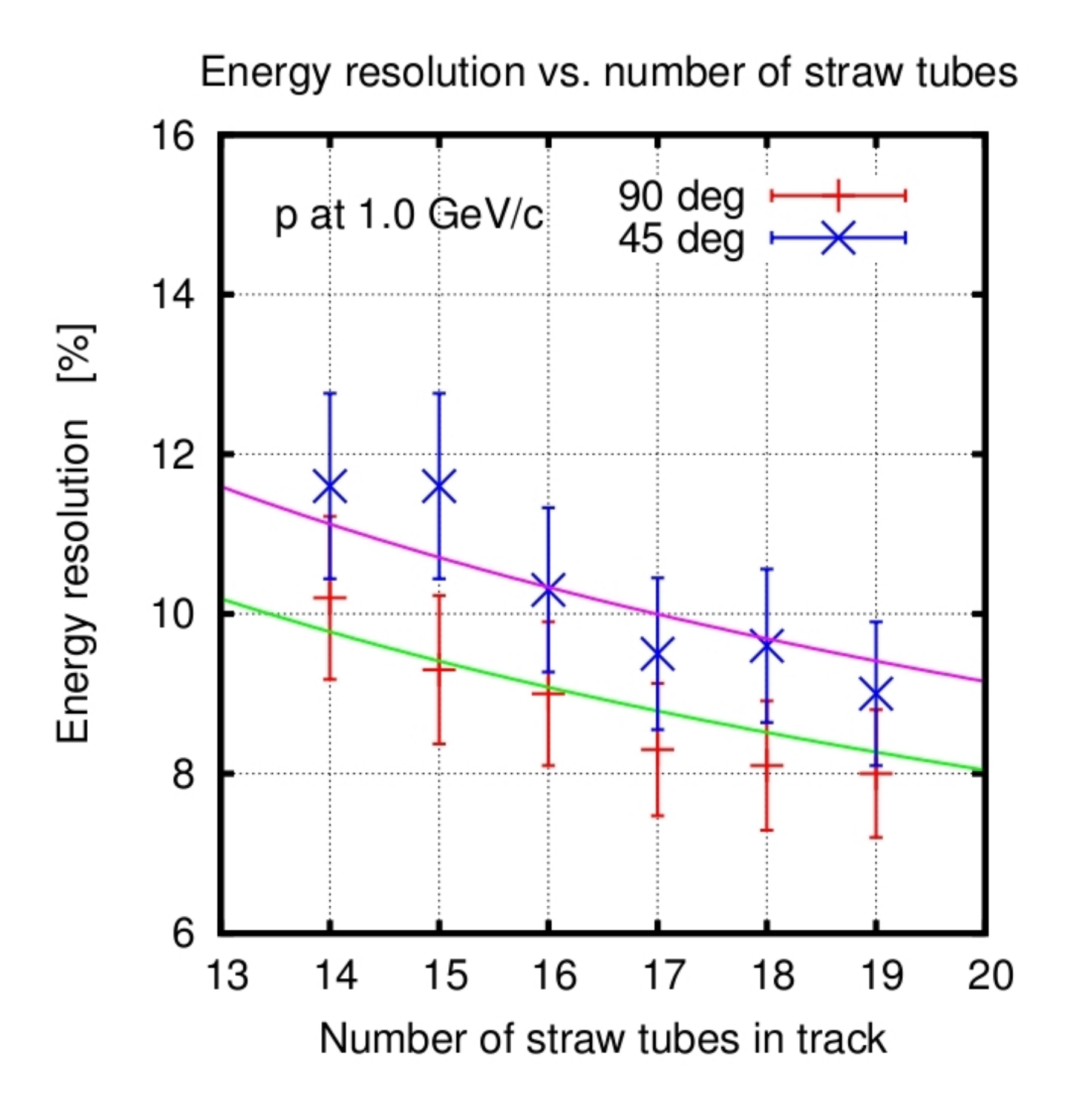}
\caption{Dependence of the energy resolution on the number of hit straws for protons with 1.0 GeV/$c$ momentum. The setup
has been inclined of 45$^{\circ}$ with respect to the beam direction.}
\label{fig:dedx:ang}
\end{center}
\end{figure}

A  set of measurements has also been performed varying the proton's impinging angle on the detector. 
For tracks inclined of 45$^{\circ}$, a systematical deterioration of the resolution of about 1\% is observed, while no significant effect on the energy resolution is observed for tracks hitting the straws at different longitudinal positions.

\begin{figure}
\begin{center}
\includegraphics[width=\columnwidth]{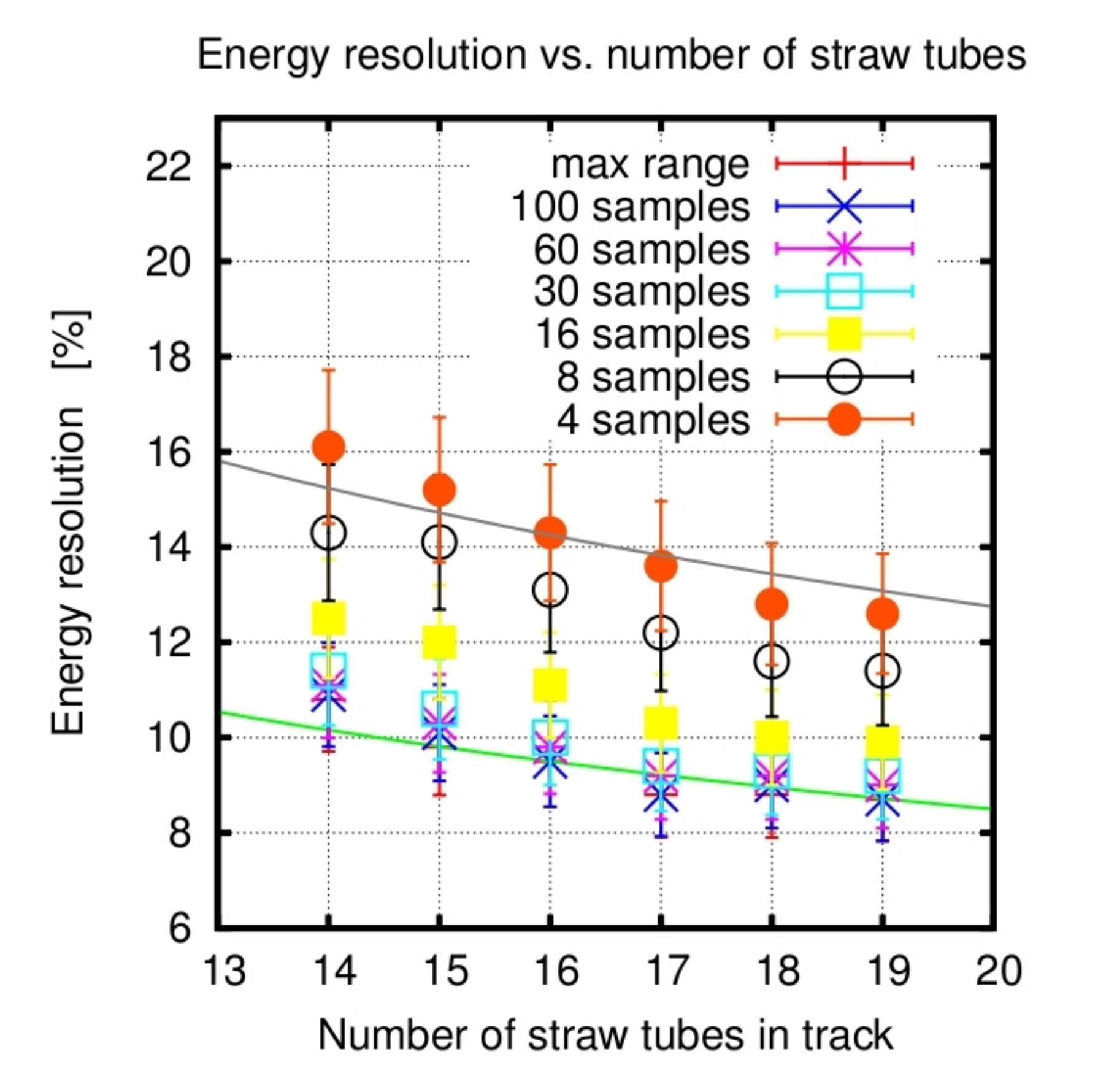}
\caption{Energy resolution obtained by decreasing the fraction of the integrated charge. The superimposed curves are functions $\propto (n)^{-1/2}$, where $n$ is the number of hits.}
\label{fig:dedx:frac}
\end{center}
\end{figure}

In order to check the possibility to perform dE/dx measurements for high particle rates, which are expected in the innermost layers of the \PANDA central tracker,  we explored the possibility of shortening the signal integration time. This will be necessary also if we will decide to use the same signal both for timing measurements and for energy-loss determination by means of the Time-Over-Threshold method.
By using the data collected for protons of 1.0 GeV/$c$ momentum, an analysis changing the fraction of the integrated signals has been performed.
The signals have been integrated over 4, 8, 16, 30, 60 and 100 flashADC samples (sample width is 4.17 ns). A significant deterioration of the energy resolution is observed when the integration range is reduced to less than 16 samples (time interval 164.17 ns). The indication ``max range'', in Fig. \ref{fig:dedx:frac}, means that the integration is done over the whole time window of the flashADC.
From simulations we know that for a 1 GeV/$c$ track, the central straw tube detector allows about 25 energy loss measurements, 
then the extrapolation of the test results demonstrates that  the needed energy resolution can be reached  in the experiment even integrating the straw signals over a time window of width $\sim$ 40 ns \cite{tdr}. 

\section{Electronics development}
\begin{figure}[!h]
\begin{center}
\includegraphics[width=\columnwidth]{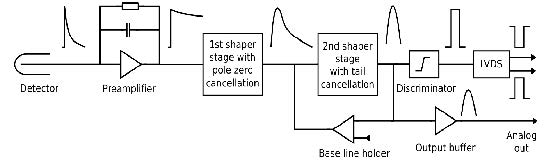}
\caption{Block diagram of the new ASIC developed for the straw tube readout.}
\label{fig:asic}
\end{center}
\end{figure}

The electronic chain used for the prototype tests has been chosen to get the best results from the detector in term of spatial 
and energy resolution, but could not be the final one. 
Space and cost constraints, combined with the need to be integrated with the \PANDA DAQ general architecture, impose to design a 
different readout scheme.
To fulfill these requirements, the proposed straw tube readout comprises three stages:
\begin{itemize}
\item Analog front end electronics cards placed close to the detector. These will host high voltage distribution, 
a signal amplification stage, a shaping circuit, and a discriminator unit with differential outputs;
\item Digital Boards (DB), for time and amplitude (or charge) measurements, located outside the spectrometers in a 
region without magnetic field and with a low level of radiation. These will comprise local logic resources for
noise suppression, fast hit detection, memory buffer for hit storage, serial Gbit optical links
for the data transmission and slow control;
\item Detector Concentrator Boards (DCB) receiving and merging inputs from several DB in a local memory 
buffer and sending it to the \PANDA DAQ system.
\end{itemize}
For the first stage, a new ASIC is under development. It is based on AMS CMOS 350 nm technology and the first realized
prototype houses four channels. Each channel includes a charge preamplifier, a pole-zero cancellation network, 
a shaper stage, a tail cancellation network, a leading-edge discriminator, and a baseline holder. 
The chip provides differential LVDS signals for the discriminated outputs, and an analog output pulse extracted after 
the first shaping stage. The block diagram of a single channel is shown in Fig.\ref{fig:asic}
\begin{figure}[!h]
\begin{center}
\includegraphics[width=\columnwidth]{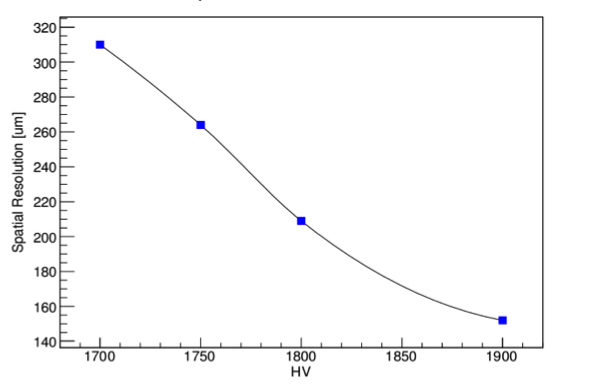}
\caption{Mean spatial resolution obtained with the new electronic chain for different high voltage settings.}
\label{fig:newASIC}
\end{center}
\end{figure}
\par
For the Digital and the Concentrator boards, their functionalities can be combined in a single Trigger Readout Board (TRB) \cite{trb3}. 
This module is the evolution of a multi-purpose device developed for the HADES data acquisition, and contains a single computer chip 
running Linux. 
\par
The new version includes also a 2 Gbit/s optical link and an interface connector (15 Gbit/s) in order to realize an 
add-on card concept which makes the hardware very flexible.
Moreover, a FPGA chip (Xilinx, Virtex 4 LX 40) and a TigerSharc DSP will provide new computing resources which can be used to run on-line analysis algorithms. 
\begin{figure*}[!t]
\centering
\subfloat{\includegraphics[width=\columnwidth]{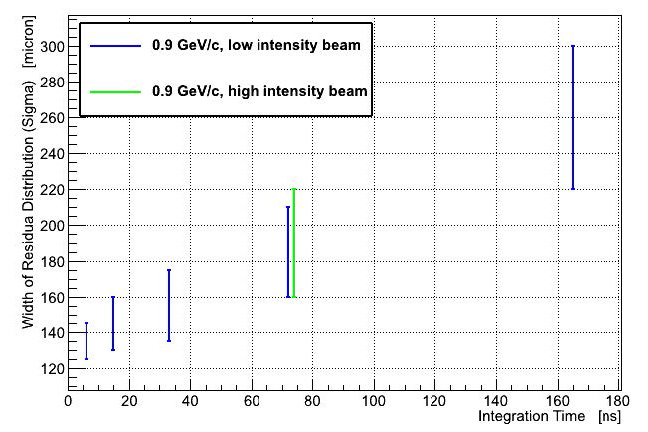}}
\hfil
\subfloat{\includegraphics[width=\columnwidth]{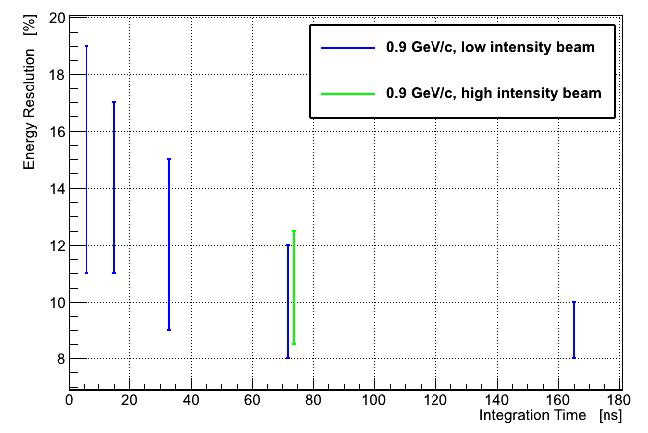}}
\caption{Test beam results for spatial (left), and energy (right) resolution as a function of the signal integration time.}
\label{fig:testres}
\end{figure*} 
\par
This readout scheme has to assure the same performance, in terms of spatial and of energy resolutions, of the test setup. 
Therefore, a straw tube prototype of 32 elements, arranged in a double-layer, has been instrumented with the new electronic chain, 
and it is presently under test.
The first measurements, performed to check the tracking capabilities, look promising (see Fig. \ref{fig:newASIC}).
\par
An other aspect that presently is under study, is the possibility to perform energy measurements by means of the Time-Over-Threshold (TOT) technique \cite{tdr}.
To extract the energy-loss information from the measured TOT, it is necessary a quite long integration time. 
This could be somehow in contrast with the request to have a short picking time which ensures a good spatial resolution and a 
reduced dead-time for the detector. On the other hand, if this approach will be successful, the DBs will require only TDC devices allowing
to reduce the amount of cables, and the costs of the electronics.
The first step of this analysis, has been to determine an appropriate signal integration time allowing to get both the required spatial 
and energy resolution. 
Fig. \ref{fig:testres} shows that for signal integration times between 40 to 60 ns, this can be feasible. Nevertheless, big spreads of threshold 
level and of charge gain among different ASIC channels, have limited the TOT capabilities of the new setup. Hence, a new, and improved version 
of the ASIC is under preparation.

\section{Conclusion}
Self-supporting straw tube modules are a good solution to reduce the overall material budget of the \PANDA tracking
systems. 
Besides the usual advantages of a conventional straw tube tracker, the \PANDA setup has many extra features:
\begin{itemize}
\item the mechanical stability of these thin-wall detectors is provided simply by the gas over-pressure inside the straw tubes 
allowing to avoid heavy mechanical frames;
\item first measurements of tubes performances in term of spatial resolution have proven that the adopted solution provides  
a spatial resolution of about 150 $\mu$m that matches the requirements of the \PANDA experiment keeping costs and detector's 
complexity under control;
\item the possibility to perform particle identification by using straw tube signals has been also exploited with promising results.
\end{itemize}
\par
This work has been partially supported by the EC Integrated Activity of FP7, HadronPhysics3, contract n. 283286.

\newpage

\end{document}